# Scaling of the minimal step-step distance with the step-bunch size:

*Theoretical predictions and experimental findings*

Katarzyna Siewierska[1], Vesselin Tonchev[2,3]

[1]University of Dublin, Trinity College, Dublin, Ireland

[2] Rostislaw Kaischew Institute of Physical Chemistry – Bulgarian Academy of Sciences, Sofia, Bulgaria

[3]Faculty of Mathematics and Natural Sciences, Card. S. Wyszynski University, Warsaw, Poland

**Abstract.** We review the studies on the scaling of the minimal step-step distance $l_{\min}$ in the bunch with the bunch size $N$, $l_{\min} \sim N^{-\gamma}$. We build our retrospective around the different values of the exponent $\gamma$ obtained from models and experiments. It was mainly studied in the context of the electromigration driven instability on Si(111) vicinal surfaces. In this context the full scaling relation is given in general form as $l_{\min} \sim \left(A/l_c FN^2\right)^{1/(n+1)}$, where $A$ is the magnitude of the step-step repulsions with range $n$, $F$ is the electromigration force acting on the charged surface atoms and $l_c$ is a length-scale, characteristic for the regime of step bunching, diffusion-limited (DL) or attachment-detachment limited (KL).

**Introduction**

The term *step bunching* (SB) refers to a broad class of surface instabilities. In the process of their motion, mediating growth or evaporation/dissolution, the steps from a vicinal crystal surface lose the initial equidistant distribution to form groups of *straight* steps called *step bunches*. Thus on the crystal surface remain wide areas almost free of steps, usually called *terraces,* although in a more strict context *terrace* is called also the distance between two neighboring steps.

Step bunching on vicinal crystal surfaces is among the most studied routes to surface patterning, both experimentally [1-15] and theoretically [16-33], to mention here only a few of them, and the development in the field was timely reviewed[34,35]. Nowadays a variety of real-world studies are reported on how to use self-organized surface patterning for concrete bottom-up applications [36-44]. In the *pre-modern* times often studied was the electric current effect on tungsten surfaces [45-48] mostly because the ubiquitous use of incandescent lamps provided abundant quantities of experimental samples and reorganization of surface structure was reported, but the term *step bunching* was not used to describe the resulting morphologies.

In a recent *classification of SB phenomena* [49] two types of step bunches are identified according to the behavior of the minimal step-step distance in the bunch $l_{\min}$ with increasing the bunch size *N*. In the B1-type $l_{\min}$ is constant as *N* increases , while in the B2-type $l_{\min}$ decreases with the increase of *N*. The quantification of the latter behavior is the focus of our mini-review. The number in the notation of the bunch type gives the number of the length-scales necessary to describe the evolving surface morphology – 1 (bunch height/size *or* bunch width) or 2 (bunch size *and* bunch width). Within the B2-type the SB phenomena are divided further into *universality classes* [25]. In those classes the two length-scales, the bunch width *W and* the bunch size *N*, are distinguishable as quantified by the corresponding time-scalings. The *protocol for SB studies* [29] is designed to study SB in models and experiments from a unified perspective. It contains two monitoring schemes based on important definition - *bunch distance* is any distance between two nearest steps smaller than the initial (vicinal) one. The *bunch* is defined as a sequence of bunch distances bound by two distances that are not bunch ones. The first of the monitoring schemes, MSI, has no memory. It acts only in a concrete moment and counts the number of bunch distances and the number of bunches. Then calculates the average bunch width, average terrace width, globally minimal step-step distance, etc., stores these data vs. the time and "forgets". The information is used also in monitoring scheme II (MSII) which is spanning the whole calculation keeping separate record on any bunch size that is observed. In this record are kept averages of the bunch width, minimal, first and last bunch distances. MSII stores what is collected from the beginning of the calculation up to the moment of writing. Another way of thinking of the number in the bunch type notation is that the B1-type needs only MSI for its study while B2 requires both MSI and MSII. For studying the scaling of $l_{\min}$ with *N* MSII is used [4,13,26,29]. At the same time, the understanding of the necessity to determine *simultaneously* the time-scaling behavior of the two length-scales within the B2-type of SB is still not a standard [49].

Most of the quantitative experimental studies on the SB instability caused by DC electric field revolve around the evaporating Si(111) vicinal surfaces. SB on Si(111) was reported for first time in 1989 by Latyshev et al. [1]. There are four step bunching regimes observed on Si(111) – two with step-down direction of the current and two with step-up direction [5] and there are further differences depending on the background physical process – (net) growth or evaporation [5]. There is also experimental possibility to compensate the evaporation and thus to observe a stationary pattern formation [50]. The vicinal surfaces of tungsten are also intensively studied [46-48]. Following the experiments of Zakurdaev [48], the first theoretical explanation of the electric field effect on a high-symmetry crystal surface is due to Geguzin and Kaganovski [51]. Their linear stability analysis identified as source of destabilization the increase of surface diffusivity with the change in crystal orientation.

The first modern theory on how the electromigration of the adatoms causes the surface steps to break their equidistant vicinal distribution was proposed by Stoyan Stoyanov [16,17] and developed further in a series of works [18-20,50,52-54]. Stoyanov identified as a source of the instability the biased diffusion of the adatoms. It makes the contributions of the both terraces bounding a step uneven. In the first version of the model the destabilizing effect of the electric field was not opposed by the step-step repulsions [55,56]. They were incorporated later in the step dynamics by Natori [57].

In what follows we review the various results for the scaling relation $l_{\min} \sim N^{-\gamma}$, ordered by the values of $\gamma$. These are the scaling relations that help to quantify the observations and thus to navigate through the complex phenomenon. However, the overall physical picture is still largely missing.

## II. Theoretical and experimental results for $\gamma = 2/3$

Three theoretical studies reveal the different regimes of the instability. In terms of the kinetic length $d \equiv D_s/K$, where $D_s$ is the surface diffusion coefficient of the adatoms and $K$ is the kinetic coefficient of adatom-to-step attachment/detachment, the regimes are $d=a$, where $a$ is the size of the lattice site, $d<<1$ and $d>>1$. The first of these is, in fact, a special case of the DL-regime. All three studies are based on the notion of *non-transparent* (*impermeable*, *opaque*) steps. *Non-transparency* is a feature of the surface steps that does not allow the diffusing adatoms to cross them without attaching to the crystal phase at kink position. In the case of *transparent* steps, the adatoms can cross the steps without attaching to a kink. In terms of activity non-transparent steps exchange actively adatoms with the terraces and it is more likely to observe DL regime as a manifestation of the step non-transparency. The idea of step transparency is easier to be understood in the context of the surface evaporation – the adatoms do not feel the steps while diffusing but still they are emitted from the steps[20]. It is the step non-transparency that permits to study the vicinal surfaces in the approximation of Burton, Cabrera and Frank (BCF) [58], in which the considerations of the adatom concentration are restricted on a single terrace bounded by two immobile steps. The general idea is to obtain the velocity of the steps from the gradients of the concentration at the steps in a self-consistent way.

II.1 First report on the size-scaling of $l_{\min}$ [19] with $d = a$

Stoyanov and Tonchev reported for first time [19] the analytical and numerical finding of scaling relation between a distance in the bunch and the number of (non-transparent) steps $N$ in it. Here we will sketch their derivation. Starting point of the study [19,54] is the model proposed by Stoyanov yet in 1991[17] - a stationary diffusion equation containing three terms (notations are slightly changed):

$$D_s \frac{d^2C}{dx^2} - \frac{D_s F}{kT}\frac{dC}{dx} - \frac{C}{\tau} = 0 \qquad (1)$$

where $C$ is the surface concentration of the adatoms, $F$ is the electromigration force and $\tau$ is the average time the adatoms spend on the surface after detaching from the steps and before evaporating in the ambience. The ratio $D_s F/kT$ is called *drift velocity* and it is a measure of how the electric field biases the diffusion field of the charged adatoms. In the boundary conditions on the two steps bounding the terrace are equated the fluxes in these two points (contribution of this terrace) to the velocity of step motion, being equal to the product of $K$ and the effective under-saturation at that point. The effect of step-step repulsion of the form $U = A/\Delta x^n$ with canonical range of interactions $n=2$[56], for two steps a distance $\Delta x$ apart, is taken into account through its influence on the equilibrium concentration at the step. This is an important improvement of the original model [17] – without step-step repulsion the development of the instability would lead to the formation of macrosteps because nothing opposes such a formation, as shown in an early computational study by Grishtenko and Lantsberg [59]. Further, the velocity of step motion is obtained as a sum of the two fluxes from the two terraces surrounding the step. They are given (equation 10 in [19]) in a dimensionless form as the velocity of *i*-th step,

$$\frac{dx_i}{dt} = f(P_i, \Delta x_i, \Delta x_{i+1}, \Delta x_{i-1}, \Delta x_{i+2}) \qquad (2)$$

which is function of four step-step distances of the form $\Delta x_i = x_i - x_{i-1}$ - the widths of the two nearest neighboring terraces, and the widths of the two next-nearest neighboring terraces, containing also four parameters $P_i$. For the non-dimensionalization procedure resulting in eq. (2) the initial vicinal distance $l_0$ was used as a length-scale and $l_0/K$ was used as a time-scale. The dimensionless parameters are $P_1 \equiv Fl_0/2kT$ reflecting the destabilizing role of the electromigration force $F$ (the surface described by this model is unstable only when the electromigration force is pointing step-down) and $P_2 \equiv l_0^{-1}(na^2 A/kT)^{1/(n+1)}$ accounting for the stabilizing role of the step-step repulsions. The other two parameters contain the mean diffusion distance $\lambda_s \equiv \sqrt{D_s \tau}$. These are $P_3 \equiv l_0/\lambda_s$ and $P_4 \equiv \lambda_s/a$. They determine the rate of vicinal motion. The analytical scaling is obtained for the *average* step-step distance in the bunch $l_b$ [19] from integrating a continuum equation for the evolution of the surface height, derived from the equations of step motion, eq. (2):

$$l_b = \left(11.45 \frac{aA}{|F|N^2}\right)^{1/3} \qquad (3)$$

The full scaling relation contains a combination of the model parameters. This is a demonstration of the idea of *scaling*, called also *scale invariance* – the individual parameters determine the system's behavior only in specific combination called *scaling parameter*. In the concrete case for a characteristic length $l$, minimal or average one, it is :

$$l/l_0 \sim P_2 \left(P_3 P_4 / P_1\right)^{1/3} \qquad (4)$$

Provided the scaling parameter, eq. (4), has the same numerical value, one cannot distinguish in between the different values of the parameters leading to the same result. In other words, the system's behavior is *universal* with respect to the property $l$. In the regime of intermediate asymptotics [60], some details can disappear from the overall picture thus proving unimportant. Here it is the mean diffusion distance $\lambda_s$ and what is important for the determination of $l$ is only the balance between the stabilizing and destabilizing tendencies, quantified by $P_2$ and $P_1$, respectively.

What is highly non-trivial in the scaling relation above, eq. (3), is the dependence on $N$. It is not obvious how it could be obtained by a dimensional analysis although such examples are available [60]. Only the numerical integration of the original system of equations, eq. (2), combined with an adequate monitoring protocol can further provide an important information. First, to check the validity of the analytical expressions derived by some inevitable approximations. Second, to determine the values of the numerical pre-factors (*amplitudes*) in the scaling relations. We illustrate this by digitizing the data points from fig. 4 in [19] for the dependence of $l_{min}$ on the bunch size $N$ obtained from numerical studies using *isolated* (single) bunches – one prepares a bunch with given size and the integration of the equations of step motion continues until a stationary value of $l_{min}$ is achieved. We see in Figure 1 that when combining the parameters used to obtain these points [19] according to eq. (4) and fitting the numerical pre-factor in order to have a best fit through the points one obtains:

$$l_{min} = \left(\frac{8aA}{|F|N^2}\right)^{1/3} \qquad (5)$$

Thus, the analytical result for $l_b$, and the numerical result for $l_{min}$ differ by a factor of 1.13 [19]. The exact form of the scaling, eq. (5), was re-derived in [26], eq. 65 therein. We illustrate further in Figure 1 the scaling obtained from the model in [19] when studying systems of many steps equally spaced in the beginning. In Figure 2 are shown the typical for the B2-type SB surface profile and surface slope obtained in these calculations.

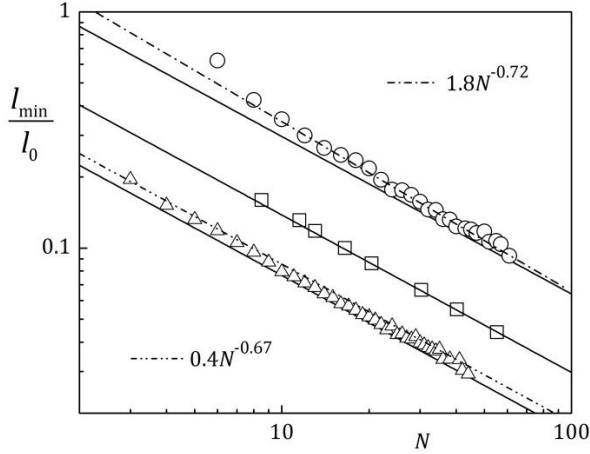 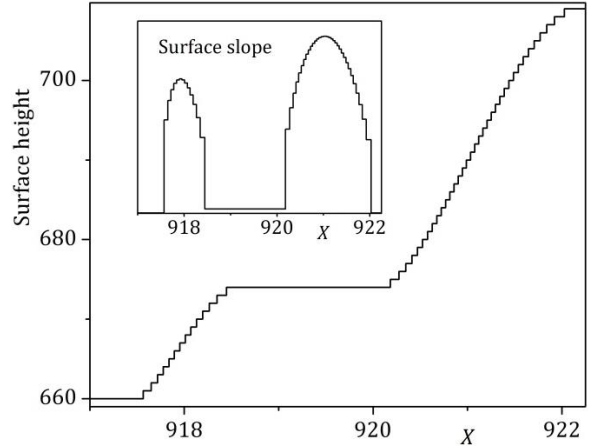

Figure 1 The digitized data (squares) from fig. 4 in [19], parameters are $P=(-0.00006, 0.003, 0.0025, 117000)$. The other data sets are obtained from the same model by using MSII with slightly different than the original parameters: $20P_1$ and $5.83P_2$ (circles), and $0.1709P_4$ (triangles). The straight lines correspond to eq. (5).

Figure 2 Surface profile and surface slope obtained by the model in [19], the data marked with triangles in Fig. 1, parameters are $P=(-0.00006, 0.003, 0.0025, 20000)$. In these calculations are used 1000 steps, starting with an almost equidistant step distribution.

Soon after [19] their scaling relation was used in the context of nanostructure growth on step-bunched Si(111) surfaces [36].

II. 2 The size-scaling of $l_{min}$ in the DL regime, $d \ll 1$, obtained by Sato and Uwaha [22]

Sato and Uwaha studied the diffusion-limited (DL) regime of the instability by adding the effect of the step-step repulsions to the first Stoyanov's model [16] where no surface evaporation is present, and obtained a slightly different from (5) scaling relation:

$$l_{min} = \left( \frac{8aA}{FN^2} \frac{a}{d} \right)^{\frac{1}{n+1}} \qquad (6)$$

In it enters also the ratio $d/a$. Note that eq. (6) reduces to eq. (5) when $d = a$.

This model was studied further by Chang et al. [61] and Ranguelov et al. [62] and, besides obtaining regimes with $\gamma = 2/3$, they find from the same model other regime(s) with $\gamma = 1/2$. We return to this model later and obtain results from it in Section IV.

III. 3 Four rounds of experimental confirmations: $\gamma \approx 2/3$

In two independent studies, Fujita et al. [63] and Homma and Aizawa [6], both on evaporating vicinal Si(111), confirmed the predicted value of the size-scaling exponent, $\gamma = 2/3$, for the case of SB due to heating with DC directed across the steps down, the same is the direction of diffusional bias for the Si-adatoms. At that time, the value of 2/3 was considered decisive for the determination of the step-step repulsion range as $n=2$.

Instead of studying the size-scaling of $l_{min}$, Fujita et al. [63] studied the maximum slope $m_{max}$ in the step-bunched regions and plotted it against the bunch height $H$. Both properties, slope and height, can be easily transformed into the quantities used above when having the value $h_0$ of the atomic height on the Si(111) surface, it is 0.314 nm [14]. Using this, one obtains $l_{min} = h_0/m_{max}$ and $N = H/h_0$. Fujita et al. [63] found $m_{max} \sim H^{0.69}$ at 1250°C. Homma and Aizawa [6] studied the average, instead of the minimal, step-step distance $l_b$ in the bunch as function of its size $N$ and they found $\gamma \approx 0.7$ for the two temperature intervals where SB is observed with step-down current - 860 °C –960 °C and 1210 °C –1300 °C. These systematic deviations of $\gamma$ towards higher values in both studies remain an open question. Note that also Popkov and Krug [27] obtained the same value within an analytical treatment. Here we provide a qualitative discussion based on Figure 1. The two data sets are obtained within the "natural bunching geometry" [26] using the monitoring scheme II (MSII) to study the time evolution of a system with 1000 steps. They show bigger deviations up from the $N^{-2/3}$ lines, eq. (5), for the smaller values of $N$. This "correction to scaling" effect is more pronounced for the larger value of the parameter combination (circles), eq. (4). In other words, when the destabilization effect is smaller the predicted scaling regime is achieved for larger bunch sizes/times and this could contribute to the observation of size-scaling exponents that are fictitious while the true scaling regime is always the same. Note that it is in principle hard to control in experiments the parameters separately and this is particularly true in the experiments on Si(111) – by fixing the value of the direct current both the temperature and electromigration force are fixed. Thus the only other controllable parameter is the vicinal miscut angle that fixes $l_0$ and another set of experiments was designed with the idea to vary the miscut while keeping all other parameters the same [14]. It was only recently when a sophisticated equipment was developed to decouple the temperature and the strength of the electric field [13]. Both these sets of experiments [13,14] were measuring also the value of $\gamma$ in different temperature intervals. Gibbons et al. found [14] $\gamma=0.70$ with step-up current (1090°C) and $\gamma=0.71$ with step-down current (1290°C). Usov et al. measured [13] $\gamma=0.64-0.65$ at 1270°C. Last results were obtained [13] at a fixed temperature but with varying the electric fields 2 to 4 times.

II. 4. Krug et al. [26] revisit the KL model of Liu and Weks[21], $d \gg 1$

The study [26] is based on the most studied model in the field of SB – this of Liu and Weeks (LW) [21]. Almost in parallel with the study already discussed [19] Liu and Weeks [21] introduced this simple model with parameters that are tuned to be relevant for the Si(111) at 900 °C. The model is derived for the kinetics-limited regime and given in a universal (dimensionless) form[64] as:

$$\frac{dX_i}{dT} = (\Delta X_i + B\Delta X_{i+1}) + (2F_i - F_{i+1} - F_{i-1}) \quad (7)$$

with $F_i \equiv \Delta X_i^{-(n+1)} - \Delta X_{i+1}^{-(n+1)}$, the dimensionless distance between two neighboring steps being $\Delta X_i \equiv X_i - X_{i-1}$. The initial vicinal distance $L_0$ is the same for all nearest step-step distances. In this version of model definition $L_0 \equiv l_0 / (nAKC_0 a^4 \tau / kT)^{1/(n+2)}$ is dimensionless and $B \equiv (1-b)/(1+b); b \equiv -KC_0 a^2 F\tau / kT$. Although we give here the notations for the case of electromigration affected sublimation, the model is valid also for the other kinetic-limited instabilities – evaporation with normal Ehrlich-Schwoebel (ES) effect and growth with inverse ES effect. When $B < 1$ this means that the contribution of $\Delta X_i$, the terrace behind the step, to its velocity is bigger and the vicinal surface is unstable. Larger values of $L_0$ only amplify the instability. The second term in eq.(7) reflects the stabilizing role of the step-step repulsion. Liu and Weeks focused their study [21] mainly on the time evolution of the bunch size $N$, while the need of monitoring in parallel the bunch width $W$ [25,29,49] was still not recognized at that time.

Krug et al. [26] obtained the full scaling relation for the kinetics-limited regime of electromigration-affected vicinal sublimation as:

$$l_{min} = \left( \frac{8aA}{|F|N^2} \frac{2a}{l_0} \right)^{\frac{1}{n+1}} \quad (8)$$

Note that here, instead of $d/a$ in the former case, eq. (6), enters the initial vicinal distance as the ratio $l_0/2a$. This scaling relation was derived implicitly by Liu and Weeks [21]. In [26] it was obtained both theoretically and numerically and the results from this study, a complete theory of step bunching in 1D, could serve as reference in the field of SB studies.

II. 5 Gibbons et al. [14] check experimentally the regime: diffusion-limited or kinetics-limited?

Now it is tempting to use these two predictions, eqs. (6) and (8), to try to distinguish in between the two regimes, KL and DL, experimenting with a real system such as the most studied one – evaporating vicinal Si(111) surfaces. As already mentioned the only other controllable in the experiments parameter besides the electric field is the initial vicinal distance $l_0$. Gibbons et al. [14] used the fact that the product $l_{min}N^{2/3}$ depends on $l_0$ only in the KL regime, eq.(8). They prepared the Si(111) samples with special grooves on them, in order to have a range of initial vicinal distances $l_0$, and studied the dependence of the product $l_{min}N^{2/3}$ on $l_0$ at three temperatures, at two of them - 940 °C and 1290 °C, the SB is with step-down direction of the current and at one - 1090°C, with step-up. What is not clear is that the measured values of $\gamma$ are 0.49 (940°C), 0.70 (1090°C) and 0.71 (1290°C) and only the value obtained at 1290°C is expected based on the predictions and previous experiments. Gibbons et al. [14] have shown that $l_{min}N^{2/3}$ does not depend on $l_0$ at the three studied temperatures and concluded that the SB regime is always diffusion-limited.

## III. Theory and experiments with $\gamma = 3/5$

Stoyanov [18,20] found an original way to treat the problem of step bunching with the step-up current direction based on the hypothesis of step transparency. Since the steps are considered transparent in this case, the BCF theory is no more applicable because of the strong coupling of the diffusion fields on the neighboring terraces. Stoyanov [20] treats the step bunch within a continuum model where an important simplification is achieved by using in advance the slope of the bunch to be obtained (in a non-self-consistent manner). For the far-from-equilibrium process, Stoyanov [20] obtained:

$$l_b = \left( 155.63 \frac{a^2 A \lambda_s}{F} \frac{1}{N^3} \right)^{\frac{1}{n+3}} \tag{9}$$

For the near-to-equilibrium case Stoyanov predicted [20] $l_b \sim N^{n/(n+2)}$.

The experimental confirmations are ordered in the table at the end. Note that Usov et al. [13] checked also the dependence of $l_{min}$ on the electric field and for the step-up case they were not able to confirm the prediction of $l_{min} \sim E^{-1/5}$ as follows from eq. (9) with $n=2$. Instead, they found for both step-up and step-down current regimes $l_{min} \sim E^{-1/3}$.

## IV. Open discussion

In this section, we do not provide answers but try to ask the proper questions. The eventual answers could possibly contribute to the still missing overall physical picture of the SB on Si(111) having four temperature intervals where the different current directions causing the phenomenon alternate. We have in mind also unpublished experimental results on the observation of SB caused by electric field on metal surfaces [65].

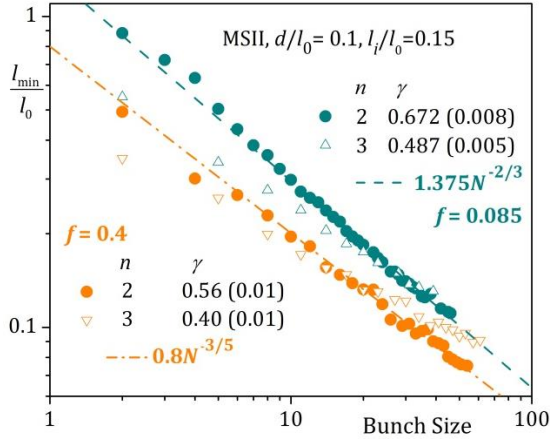
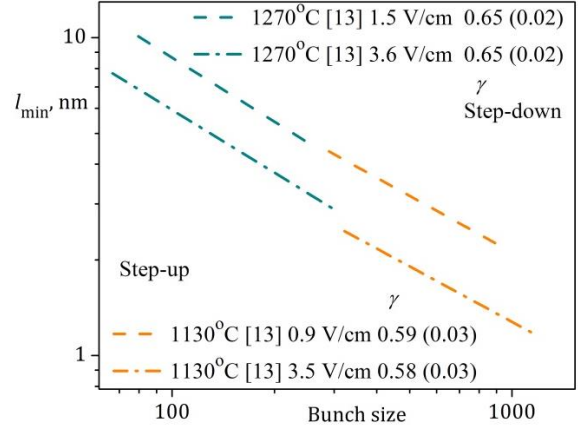

Figure 3 Numerical calculations using the model of Stoyanov [16] with step-step repulsions incorporated [22] for two different values of the dimensionless electromigration force $f$ =0.085, 0.4 in the DL-regime, $d/l_0$=0.1. Dimensionless magnitude of step-step repulsions is the same, $l_i/l_0$=0.15. The scaling, eq. (6), is not plotted but can be easily calculated.

Figure 4 Semi-quantitative illustration of the experiments of Usov et al. [13]. Only the fitting lines are taken from their plots spanned by the corresponding initial and final data points. Note also that the original "slope vs. height" plots are transformed to "$l_{min}$ vs. $N$" plots by the procedure suggested in [14].

In figure 3 we plot results for $l_{min}$ vs. $N$ obtained from simplest model that results in SB with step-down electromigration and without growth or evaporation fluxes [16,22]. We show that the increasing of the (dimensionless) electromigration force $f$ could change the measured value of $\gamma$. It is, for step-step repulsions range $n$=2, 0.672 ($f$ = 0.085) and 0.56 ($f$ =0.4). This effect is not observed in figure 4 where the experiments of Usov et al. [13] are illustrated. Instead, with the increase of the electric field, hence the electromigration force, the values of $\gamma$ at given temperature are almost constant. But, for practically the same values of the fields, the slopes of the bunches at 1130°C, where SB with step-up current is observed, are higher than those obtained with step-down current at 1270°C . Towards the explanation of this effect one can use the scaling relations, eqs. (6), (8) and (9), but we leave this discussion open for now.

### V. Experiment and theory with $\gamma$ =1/3

In a study devoted to the surface instabilities during homo-epitaxy on Cu-vicinals for the case of simultaneous step bunching *and* step meandering (SB+SM) $\gamma$=0.29±0.05 was obtained by Néel et al. [66,67]. The phenomenon of SB+SM was observed also on semiconductor surfaces – SiC [7] and Si(111) [15], the latter at temperatures below the reconstruction transition from (7x7) to (1x1). The scenario of these instabilities is somewhat different – while on the Cu-vicinals at lower temperatures only step meandering is observed, on Si(111) at lower temperatures two types of SB are observed[15]. Then, in both cases, the increase of the temperature leads to SB+SM.

The exponent $\gamma = 1/3$ was recovered in the numerical studies of a new 1D BCF-type model, the so called 'C$^+$ - C$^-$'-model [29,68]. The source of destabilization in this are uneven reference ("equilibrium") concentrations on both sides of the step. According to the classification of SB phenomena [49] this model is of B2-type, but with the important difference that the first bunch distance is also the minimal one. Thus, the scaling of the minimal distance in the model is the same as the one of the first bunch distance in the models studied above [22,26] where $\gamma = 2/3$.

One of the predictions of this model can be easily checked by looking at the profile and slope of the bunches in the case of SB+SM. They should have the predicted steepest part in the beginning and not in the middle as in Figure 2.

## V. Concluding remarks

Our text is largely influenced not only by the studies reviewed here but also by the ongoing studies in the group of prof. Igor Shvets in Trinity College, Dublin. The main and rather striking result of these is the systematic observation of SB phenomena on both metal and insulator surfaces [65] quantified by using the scaling of the maximal bunch slope with the bunch height.

The scaling of the minimal step-step distance with the size of the bunch briefly reviewed here is so firmly established as a basic tool in the SB studies that when the minimal distance is zero it may look that there is nothing to be studied. This is the situation when in the models there is no step-step repulsion, or in experimental systems the step-step repulsion is possibly small, and then nothing prevents the steps from merging into macrosteps [59,69]. In this context the protocol for SB studies [29] is necessarily modified [70,71] in order to provide quantification of this less studied route to surface patterning.

**Table** Different values of $\gamma$, predicted theoretically and found experimentally

| $\gamma = 2/3; 2/(n+1)$, evaporation with step-down current direction | | | | |
|---|---|---|---|---|
| Models with non-transparent steps | | Experiments on Si(111) | | |
| Regime | Study | Temperature | $\gamma$ found | Study |
| $d = a$ | Stoyanov and Tonchev [19] | 1250°C | 0.69 | Fujita et al.[4] |
| DL | Sato and Uwaha [22], Sato and Uwaha for the case of vicinal growth destabilized by inverse ES effect [72], the parameters used are for the DL regime. | 910°C 1237°C 1256°C 1285°C 1343°C* | 0.70 0.70 0.66 0.71 0.64* | Homma and Aizawa [6] *step-up |
| KL | Liu and Weeks [21], Krug et al. [26], Popkov and Krug [27] find $\gamma = 0.7 \pm 0.1$ | 1090°C 1290°C | 0.70* 0.71 | Gibbons et al. [14], *step-up |
| | | 1290°C | 0.64-0.67, error ±0.03 | Usov et al. [13] dep. on field |
| $\gamma = 3/5; 3/(n+3)$, evaporation and step-up current direction | | | | |
| Model | | Experiments on Si(111) | | |
| Stoyanov [20], continuum model of far-from-equilibrium evaporation with transparent steps and step-up adatom electromigration | | Temperature | $\gamma$ found | Study |
| | | 1145 °C | 0.6 | Fujita et al.[4] |
| | | 1160 °C 1180 °C 1343 °C | 0.61 0.59 0.59 | Homma and Aizawa [6] |
| | | 1180 °C | 0.57-0.59, error ±0.03 | Usov et al. [13] dep. on field, |
| $\gamma = 1/2$ | | | | |
| Models | | Experiment on Si(111) | | |
| Stoyanov [20], continuum model of near-equilibrium evaporation with transparent steps and step-up adatom electromigration, $\gamma = n/(n+2)$ | | Temperature | $\gamma$ found | |
| | | 940°C evap. with step-down current | 0.49 | Gibbons et al. [14] |
| Krug et al.[26], analytical prediction with non-transparent steps and step-down adatom electromigration, $\gamma = 2/(n+2)$ | | | | |
| Chang et al. [61], Ranguelov et al. [62], based on the model from [16,22] - step-down adatom electromigration and no evaporation | | | | |
| $\gamma = 1/3; 1/(n+1)$ | | | | |
| Models | | Experiments on Cu homo-epitaxy | | |
| BCF-type model with non-transparent steps and destabilization from the uneven reference ("equilibrium") concentrations [29,68] – the 'C$^+$ - C$^-$'-model | | $\gamma$ found | Study | |
| | | 0.29±0.05 | Neel et al. [66,67] | |

**Acknowledgements**. VT is supported in part by the Collaborative Research Program of the Research Institute for Applied Mechanics, Kyushu University and T02-8/121214 from the Bulgarian NSF. He thanks Cormac O'Coileain, Olzat Toktarbaily and Victor Usov for the useful discussions. The authors express their sincere gratitude to Professor Igor Shvets from the Trinity College in Dublin, thanks to him they started their collaborations.